\documentclass[runningheads]{llncs}

\usepackage[T1]{fontenc}
\usepackage{graphicx}
\usepackage{todonotes}
\usepackage{float}
\usepackage{placeins}

\begin{document}

\title{Automatic Prompt Optimization Techniques: \newline Exploring the Potential for Synthetic Data Generation}
\titlerunning{Automatic Prompt Optimization Techniques}

\author{Nina Freise\inst{1,2}\thanks{These authors contributed equally.}
Marius Heitlinger\inst{1,2}*
Ruben Nuredini\inst{1}
Gerrit Meixner\inst{1}}

\authorrunning{N. Freise, M. Heitlinger et al.}
%
\institute{Heilbronn University of Applied Sciences \and
Heidelberg University}

\maketitle              
\begin{abstract}
Artificial Intelligence (AI) advancement is heavily dependent on access to large-scale, high-quality training data. However, in specialized domains such as healthcare, data acquisition faces significant constraints due to privacy regulations, ethical considerations, and limited availability. While synthetic data generation offers a promising solution, conventional approaches typically require substantial real data for training generative models. The emergence of large-scale prompt-based models presents new opportunities for synthetic data generation without direct access to protected data. However, crafting effective prompts for domain-specific data generation remains challenging, and manual prompt engineering proves insufficient for achieving output with sufficient precision and authenticity. We review recent developments in automatic prompt optimization, following PRISMA guidelines. We analyze six peer-reviewed studies published between 2020 and 2024 that focus on automatic data-free prompt optimization methods. Our analysis reveals three approaches: feedback-driven, error-based, and control-theoretic. Although all approaches demonstrate promising capabilities in prompt refinement and adaptation, our findings suggest the need for an integrated framework that combines complementary optimization techniques to enhance synthetic data generation while minimizing manual intervention. We propose future research directions toward developing robust, iterative prompt optimization frameworks capable of improving the quality of synthetic data. This advancement can be particularly crucial for sensitive fields and in specialized domains where data access is restricted, potentially transforming how we approach synthetic data generation for AI development.

\end{abstract}
\section{Introduction}

Artificial intelligence (AI) technologies are rapidly transforming diverse sectors and fundamentally reshape operational paradigms across industries. The profound impact of these systems is critically dependent on the reliability of their decision-making capabilities. This reliability is fundamentally tied to the AI-models' predictive capabilities which critically depends on two key dimensions: the comprehensive scale of training data and the complexity of model architecture. The shift toward data-centric approaches \cite{EffectivenessData} has further intensified the importance of comprehensive, high-quality input data. Modern AI systems, based on learning methods such as deep learning, rely heavily on massive datasets to learn complex patterns and representations \cite{sutton2019}. This paradigm shift has made access to comprehensive, heterogenous, and high-quality datasets a fundamental prerequisite for the development of sophisticated AI systems.

Obtaining suitable data for AI systems presents numerous challenges. These include: rarity of data samples, resource-intensive data collection processes, and inconsistent data recording protocols. Further difficulties are caused by inherent biases in the collection and labeling, the unstructured nature of raw data, as well as strict privacy regulations governing sensitive information. These challenges are particularly evident in domains where confidentiality and privacy are of paramount importance. One such is the healthcare and medical domain \cite{app13127082}. By nature, medical data is highly sensitive and personal, making its acquisition and access challenging due to ethical issues and strict privacy regulations such as GDPR or HIPAA. These traits have substantial effect on the usability of AI-powered healthcare systems as many potential users express concerns about potential privacy breaches\cite{TrustinandAcceptanceofArtificialIntelligenceApplicationsinMedicine:MixedMethodsStudy}. 

An established approach to addressing data scarcity is the use of augmentation techniques. The primary goal of data augmentation is to expand both, the volume and diversity of data samples, by leveraging a small set of real data as a basis for generating novel samples. Current approaches involve utilization of advanced deep learning architectures such as Generative Adversarial Networks (GANs) \cite{goodfellow2014}, Variational Autoencoders (VAEs) \cite{kingma2013,rezende2014}, and Diffusion Models \cite{Ho2020}. These \emph{generative models} learn to replicate the underlying statistical properties of the real data and, once trained, are a limitless source of unique \textit{synthetic} data instances. Synthetic data closely mimics real data while maintaining anonymity and privacy. Studies have highlighted the effectiveness of generative models, demonstrating that models trained solely on synthetic data can achieve reasonable performance \cite{Carrle2023}. Additionally, the combined use of synthetic and real data (a hybrid approach) further enhances the predictive performance of AI models in medical applications and often displays a superior performance over models trained solely on real data \cite{Xu2024PracticalAO,Frid_Adar_2018,Dorjsembe2024}.
One further benefit of synthetic data is that it can serve as a valuable alternative for research, teaching, and practice, free from the ethical and legal restrictions often associated with the use of real data. 

Regardless of the effectiveness, the quality of synthetic data obtained by generative models highly depends on the volume and quality of the data used for their training. This becomes challenging when extensive, high-quality datasets are not readily available, paradoxically highlighting the very issue that synthetic data generation seeks to address.

The recent emergence of the so-called \textit{prompt-based models} introduce a new paradigm for data generation. These models take a textual prompt as input and are capable to generate corresponding data as an output. Notable examples are the Large Language Models (LLMs) \cite{minaee2024} such as OpenAI's GPT \cite{openai2023gpt4} and Meta’s LLaMA \cite{meta2023llama} which are trained on vast amounts of text data, and can understand and generate human-like language. Latest breakthroughs have expanded the capabilities of similar architectures, enabling them to generate data beyond text. Multimodal transformers, for example, are now capable of producing images, audio, and video.

Data generation with prompt-based models depends on task-specific inputs, typically provided by users in the form of natural language instructions, commonly refereed to as \textit{prompts}. This poses significant limitation in the usability of these models, as the quality of the generated output is heavily influenced by the user's ability to clearly and concisely convey own intent through text. Additionally, prompt-based models are very sensitive as small changes in the input prompt can significantly influence the response. Achieving the desired outcome typically requires several manual refinements of the initial prompt until desired output is obtained. This iterative process, referred to as \textit{manual prompt crafting}, is often labor-intensive, unpredictable, and time-consuming, involving extensive trial and error.

Various strategies for crafting precise and effective inputs, known as \textit{prompt engineering techniques}, have been developed and proposed \cite{amatriain2024}. These techniques provide valuable insights and practical advices to help users effectively guide these models towards generating desired outputs \cite{schulhoff2024}. However, both, manual prompt crafting and prompt engineering techniques, demonstrate significant model-specific limitations, revealing that their effectiveness is not consistent across different models, linguistic contexts, or varying task complexities \cite{liu2024,kepel2024} challenging the assumption of universal applicability. 

Additional challenges become apparent when prompt-based models are utilized for generating highly specialized, domain-specific data. In those cases, the prompt must include extremely detailed linguistic formulation in order to capture subtle patterns in the data. Such nuanced details are very relevant in guiding the generation process but are often conditioned by complex domain knowledge or are even beyond human perception. 

Few-shot prompting \cite{FewShotLearners}—an advanced prompt engineering technique that guides models by providing strategic demonstrations within the prompt—can be considered as a potential mechanism to enhance data generation quality. This technique involves presenting the model with representative data samples and requesting it to generate similar content. However, this approach remains constrained by the fundamental limitation that real data in domain-specific scenarios is not readily available.

Previous research exploring LLMs' capabilities for synthetic medical text generation through manual prompting report significant challenges and only marginal success in producing clinically accurate and contextually relevant data \cite{uyguner2023chatgpt,Sufi2024}. The MedSyn framework \cite{Kumichev2024} presents an advanced approach to prompt crafting for the synthetic generation of clinical notes with LLMs. However, the framework relies on prior medical knowledge, represented as a Medical Knowledge Graph (MKG), that is integrated to enrich prompts for GPT-4.

While prompt-based models demonstrate significant potential for generating specialized domain-specific data, their effectiveness is constrained by the complex nuances of human-computer interaction. This limitation becomes particularly evident when direct access to real data remains unavailable. 

Optimization by PROmpting (OPRO) \cite{yang2023large} represents an innovative approach to automated prompt refinement where LLMs are used as optimization agents.
The method leverages natural language descriptions of optimization objectives to iteratively generate and refine prompt solutions. Through systematic evaluation and incorporation of successful prompts into subsequent iterations, OPRO progressively enhances instruction quality to maximize task performance and guide LLM outputs for specific applications.

The foundational principles in OPRO served as an inspiration for exploring innovative optimization techniques for automated, iterative prompt refinement. These techniques provide a promising, low-effort approach to gradually improve prompts, progressively aligning them with the user's intent. In the problem of synthetic data generation, the objective is implicitly hidden within the patterns of real data and difficult to articulate. Therefore, automated prompt optimization techniques demonstrate significant potential to enhance the usability of prompt-based models but also, extend their applicability to unorthodox domains and use cases.

We present a systematic review of prompt optimization techniques, with particular emphasis on methods suitable for generating synthetic data. Our investigation focuses on identifying sophisticated approaches that enhance data authenticity while operating autonomously and with minimal user intervention. Given the often strict privacy regulations for accessing data, such as in the medical domain, we exclusively examine methods that can function effectively without direct access to real datasets.

\section{Methods}

This review follows the Preferred Reporting Items for Systematic Reviews and Meta-Analysis (PRISMA) standards for systematic reviews \cite{liberati2009prisma}. We focus on relevant studies on the topic of automatic and iterative prompt optimization methods for text generation. We specifically aim to identify techniques capable of operating without relying on training datasets, high-performing prompt datasets, or any established form of "ground truth". Our review encompasses literature published from 2020 onwards, following GPT-2's landmark release in November 2019, which revolutionized approaches to data generation and initiated research in the field.

We conducted the search on Google Scholar, as it indexes a wide range of academic sources, including peer-reviewed journals, conference papers, and books. Furthermore, Google Scholar provides the capability to search the full text of articles, in addition to titles and abstracts, which increases the probability of locating relevant papers on the subject matter.

The search term combines \textit{Large Language Models}, \textit{automatic prompt engineering}, and \textit{text prompts} and some variations and synonyms of these topics. 

The following query is the result: 
\newline
\begin{quote}
\textit{("Large Language Models" OR "LLM") \newline
AND 
\newline
("automatic prompt engineering" OR "iterative prompt engineering")
\newline 
AND 
\newline
("text prompts")
\newline
AND 
\newline
("optimizing" OR "optimizers" OR "improved" OR "improving")} 
\end{quote}

The inclusion and exclusion criteria for the resulting papers are defined in Table \ref{tab:criteria}.

\begin{table}[t]
\centering
\caption{Inclusion and Exclusion Criteria for Literature Review}
\renewcommand{\arraystretch}{1.5} 
\setlength{\tabcolsep}{12pt} 
\begin{tabular}{|p{0.45\textwidth}|p{0.45\textwidth}|}
\hline
\textbf{Inclusion Criteria} & \textbf{Exclusion Criteria} \\ \hline
\begin{itemize}
    \item \textbf{Automatic techniques:} Contains iterative prompt engineering techniques using text prompts
    \item \textbf{Language:} Written in German or English
\end{itemize} & 
\begin{itemize}
    \item \textbf{Publication date:} Published before the year 2020
    \item \textbf{Dataset use:} Utilizes a training set or open dataset for evaluation in its method
    \item \textbf{Access:} Inaccessible without open access or not available via the University of Heilbronn or the University of Heidelberg
\end{itemize} \\ \hline
\end{tabular}
\label{tab:criteria}
\end{table}

If the papers do not contain automatic and iterative prompt engineering techniques on text prompts for generating text, they will be referenced as out-of-scope. 

The search results were exported, divided by two, and assigned to two reviewers for initial screening. Initially, records were screened based on the inclusion and exclusion criteria using the title and abstract. Subsequently, the full texts were retrieved, and a full-text screening was performed. This step involved a re-application of the inclusion and exclusion criteria. The reasons for excluding papers were documented.
To ensure the validity of each paper, we performed the full-text paper screening individually for every paper and then compared the results. The search was conducted in October 2024.

\section{Results}
\label{sec:results}

\begin{figure}[t]
\centering
\includegraphics[width=0.88\linewidth]{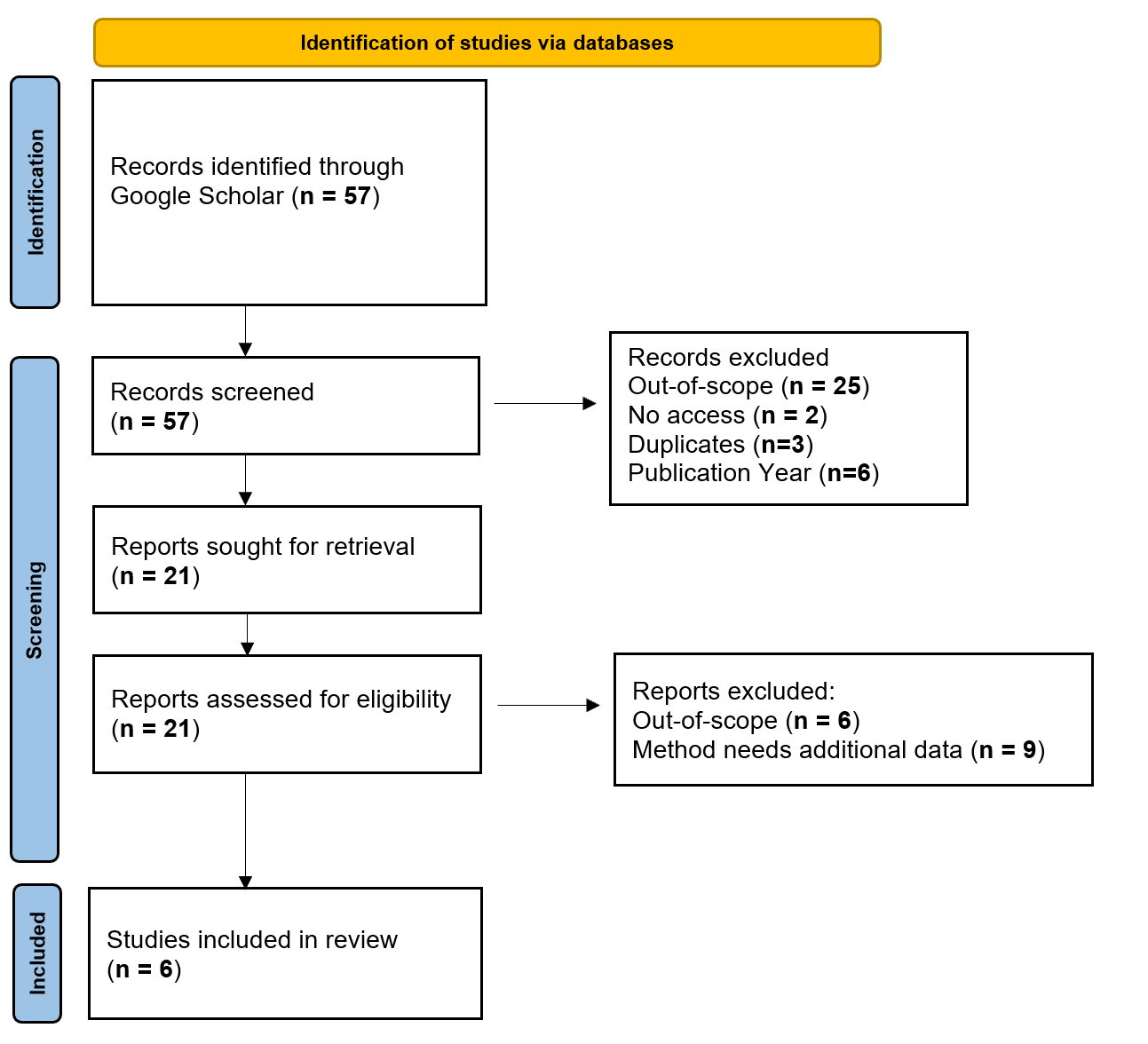}
    \caption{PRISMA flow diagram}
    \label{fig:flowdiagram}
\end{figure}

This systematic review commenced with an initial comprehensive search, which identified 57 papers. After screening the abstracts, 36 papers were excluded due to the reasons shown in Fig. \ref{fig:flowdiagram}. We subsequently conducted an in-depth analysis on 21 papers, after which we excluded 15 papers. The final count of studies considered in this systematic review is therefore 6. \\

The reviewed studies present diverse techniques for iterative prompt optimization, each leveraging unique mechanisms to refine prompt quality and model performance. Despite sharing a common objective—enhancing prompt efficacy across various applications—these approaches differ significantly in their optimization strategies, feedback structures, and scenarios where they excel.

\subsection{Feedback-Based and Iterative Refinement Approaches}

Dong et al. (2024) \cite{dong2024paceimprovingpromptactorcritic} and Wu et al. (2024) \cite{wu2024stragoharnessingstrategicguidance}  with their frameworks PACE and STRAGO adopt feedback-centric methods that enable LLMs to refine prompts iteratively, yet they differ in their feedback structures and evaluation strategies.

PACE employs an actor-critic algorithm inspired by reinforcement learning, where the LLM operates in two roles: as an "actor" that generates responses based on a prompt and as a "critic" that assesses response quality. Its feedback loop is multi-faceted; it incorporates feedback from multiple critic models per iteration, aggregating feedback to guide the optimization of each prompt. This aggregation allows for a broader perspective on prompt performance, enhancing robustness and reducing individual bias. Unlike single-step feedback loops, the actor-critic structure allows for nuanced adjustments to prompts, making it particularly effective in scenarios where varied perspectives are needed for prompt improvement. However, the model’s reliance on multiple evaluations per iteration raises computational demands, potentially limiting its applicability in resource-constrained settings.

STRAGO, while also feedback-driven, tackles a common pitfall in prompt optimization: prompt drift, where optimization for specific cases can lead to performance loss in other scenarios. To mitigate this, the method involves analyzing both successful and failed cases, offering a balanced view of prompt performance. This approach allows STRAGO to provide detailed, step-by-step refinements, addressing specific issues while preserving effective elements of prior prompts. STRAGO’s three-step process (analyzer, refiner, and optimizer) further differentiates it from PACE, with each stage focusing on dissecting both correct and incorrect predictions and generating strategies for iterative prompt refinement. By adopting this structured, multi-layered process, STRAGO is well-suited for complex, stability-dependent tasks where controlled refinement is critical, such as reasoning or domain-specific applications.

While both methods demonstrate the power of iterative feedback, PACE focuses on aggregating diverse feedback for prompt refinement, whereas STRAGO emphasizes stability by maintaining a balance between corrective actions and preserving successful prompt features. These differences highlight PACE’s strength in high-complexity environments and STRAGO’s advantage in scenarios where drift control is paramount.

\subsection{Error-Focused or Diagnostic Prompt Improvement}
In contrast to feedback aggregation, Chen et al. (2024) \cite{chen2024repromptplanningautomaticprompt} with their framework REPROMPT and Ma et al. (2024) \cite{ma2024largelanguagemodelsgood} with "Are Large Language Models Good Prompt Optimizers?" center on error diagnosis and corrective adjustments, adopting a more diagnostic approach to iterative refinement.

REPROMPT mimics a conversational feedback loop, where each response is analyzed for common points of failure. By identifying recurring errors, REPROMPT facilitates targeted prompt adjustments that address specific weaknesses rather than generic improvements. Each round refines the initial prompt, similar to a dialogue, enabling REPROMPT to iteratively shape prompts based on accumulated interaction data. This conversational structure makes REPROMPT especially effective in scenarios where task complexity increases over time, such as procedural planning or extended dialogue-based tasks. However, REPROMPT’s reliance on batch-based error analysis may limit adaptability in dynamic contexts, where feedback needs to be immediate and granular.

Ma et al. (2024) \cite{ma2024largelanguagemodelsgood} challenge the effectiveness of LLM-based reflection methods, proposing “Automatic Behavior Optimization” as an alternative. This method directly targets problematic behaviors by requiring the optimizer to identify failure steps and then refine prompts specifically at those points. Unlike REPROMPT, which iteratively refines prompts based on generalized patterns of failure, Automatic Behavior Optimization emphasizes precision by requiring the LLM to generate step-by-step instructions to address specific failures. This targeted refinement is advantageous for tasks where accuracy at each step is critical, such as instructional generation or decision-based tasks. However, the approach may be less effective in adaptive or high-variability scenarios, where failure points are not easily identifiable or repeatable.

The diagnostic focus of both REPROMPT and Automatic Behavior Optimization contrasts with the broader feedback-oriented approaches of PACE and STRAGO. REPROMPT’s conversational, batch-based error identification enables systematic adjustments across task phases, while Automatic Behavior Optimization’s step-specific refinement prioritizes direct behavioral alignment, particularly in high-precision contexts.

\subsection{Control-Theoretic and Structured Interaction Optimization}

Luo et al. (2023) \cite{luo2023promptengineeringlensoptimal} introduce a control-theoretic perspective, treating prompt engineering as a dynamic optimization problem. Unlike feedback or error-based refinement, this approach uses a structured, systematized interaction model where the initial prompt is iteratively adjusted based on response evaluation, similar to a feedback control system.

This control-based approach involves a multi-round interaction structure, where each iteration generates refined prompts to address gaps identified in previous responses. By systematically narrowing the prompt’s focus and enhancing specificity, Optimal Control’s framework excels in tasks that require controlled iterative refinement, such as diagnostic or advisory systems where multi-step accuracy is critical. However, its reliance on sequential interaction rounds can be time-intensive, potentially limiting its application in real-time scenarios where immediate responses are needed.

\subsection{Contextual and Evolutionary Optimization of Complex Prompts}
Hsieh et al. (2023) \cite{hsieh2023automaticengineeringlongprompts} employ a context-driven, evolutionary approach for prompt optimization, designed specifically for complex or lengthy prompts.

This method involves segmenting the initial long prompt into individual sentences, each rephrased independently by a mutator that maintains semantic consistency. Rather than direct feedback, the method employs a contextual bandit approach, which selects sentences likely to benefit the prompt’s overall performance when rephrased. Each revised prompt is then evaluated, and successful mutations inform future modifications, effectively evolving the prompt over successive iterations. This evolutionary approach is particularly beneficial in static, domain-specific tasks requiring detailed, multi-sentence instructions, such as legal or scientific writing. However, without direct feedback from task performance, the method may struggle to adapt to changing requirements, making it less suited for dynamic or highly adaptive applications.
\newline
\newline
These six studies illustrate a range of iterative prompt optimization techniques, each with unique mechanisms suited to different scenarios. Dong et al. (2024) \cite{dong2024paceimprovingpromptactorcritic} with PACE and Wu et al. (2024) \cite{wu2024stragoharnessingstrategicguidance} with STRAGO emphasize systematic refinement, with PACE excelling in high-complexity environments and STRAGO balancing corrective actions with stability. Chen et al. (2024) \cite{chen2024repromptplanningautomaticprompt} and Ma et al. (2024) \cite{ma2024largelanguagemodelsgood} offer diagnostic refinement strategies, prioritizing error correction for high-precision tasks, while Luo et al. (2023) \cite{luo2023promptengineeringlensoptimal} formalizes prompt engineering as a dynamic control process, beneficial for controlled, multi-round interactions. Finally, Hsieh et al. (2023) \cite{hsieh2023automaticengineeringlongprompts} leverages an evolutionary structure suited for static, multi-component prompts.

This analysis highlights the adaptability of iterative prompt optimization strategies, with each method offering tailored benefits for specific applications - feedback aggregation for comprehensive tasks, error-focused adjustments for precision tasks, structured control for interaction-heavy scenarios, and contextual evolution for complex prompts without direct feedback. 



\section{Discussion}

Our approach to systematic review was based on the PRISMA framework in order to identify relevant studies on automatic prompt engineering techniques. This approach has notable strengths, especially through its adherence to PRISMA guidelines , ensuring transparency and reproducibility. By utilizing a detailed search strategy and rigorous inclusion/exclusion criteria, the review process captured a focused set of studies that were highly relevant to the specific requirements for this study. These carefully defined selection criteria—including studies published after 2020 excluding techniques depending on explicit access to real datasets—were strategically designed to identify innovative methods for generating synthetic data. This approach is particularly crucial in domains like healthcare and medicine, where direct data access is severely restricted by privacy and regulatory constraints.

While comprehensive in scope, this systematic review has several limitations that must be acknowledged.

First, the reliance on Google Scholar, although beneficial for its comprehensive indexing, might have limited the scope to certain types of publications and excluded specialized databases, potentially overlooking niche studies. 

Furthermore, the selection criteria were restricted to publications that were either open-source or available on platforms with open-access agreements with German universities. This limitation may have excluded studies from platforms with restricted access (e.g., paywalls) or those not covered by open-access agreements, potentially omitting valuable insights for a more comprehensive understanding of the topic.

Another limitation is the language restriction of the reviewed publications as only included those written in English or German language. This may have led to exclusion of valuable studies defining effective prompt engineering techniques, potentially resulting in missed insights or innovations due to the language barrier.

Finally, the decision to exclude studies that utilized real datasets in the optimization process, while justifiable, may have omitted techniques with potentially adaptable components for iterative prompt engineering. While the exclusion criteria ensured relevance, they might have inadvertently constrained the range of methodologies available for synthesis, potentially overlooking hybrid approaches that could benefit from minor adjustments in order to be able to operate without real data.

The systematic review highlights diverse approaches to prompt optimization, emphasizing the potential for automatic, minimal-input techniques. Feedback-driven methods like PACE and STRAGO show promise in refining prompts systematically, while error-focused approaches such as REPROMPT and Automatic Behavior Optimization address specific failures for high-precision tasks. Control-theoretic methods and evolutionary strategies further diversify the landscape, each suited to unique use cases. These findings underscore the importance of balancing adaptability, precision, and computational efficiency in developing prompt engineering techniques, particularly for sensitive applications like medical data generation.

\section{Conclusion}
Prompt-based models display immense potential for generating synthetic data. However, the primary challenge lies in constructing the right prompt to invoke the generation of plausible, high-quality data — a challenge that can be addressed by automatic prompt optimization.

In this study we systematically reviewed automatic prompt optimization techniques to identify methods suitable for the task of synthetic data generation. Our focus was on approaches that do not rely on real datasets, which distinguishes them from conventional approaches in synthetic data generation by fine-tuning or training generative models. The examined methods reveal significant opportunities for further refinement and innovation, particularly in the areas of automation and user-independent design. Building on these insights, future research should aim to develop integrative frameworks that combine the strengths of different techniques into cohesive pipelines. For example, leveraging the precision of error-focused methods alongside the scalability of evolutionary strategies may yield solutions that are both efficient and robust. Such a framework would streamline the process of prompt refinement, reducing user intervention while improving the realism and applicability of generated outputs, which can play a pivotal role in addressing the growing demand for high-quality synthetic data across various domains.

While prompt engineering serves as a bridge between traditional programming (or data science) tasks and the emerging field of AI application development, its effectiveness in human-computer interaction is constrained by the challenge of accurately conveying intent. Current AI systems heavily rely on well-constructed prompts to generate relevant and accurate responses. However, this reliance is not sustainable in the long term. As AI models advance, they may develop the ability to understand tasks more intuitively, reducing the need for precise and carefully engineered prompts. Future AI systems should become more context-aware and capable of interpreting user intent with minimal or even ambiguous input. In this future scenario, automatic prompt optimization could become an essential feature of AI systems, allowing them to incorporate broader context and adapting to a wide range of interactions. Continued research and innovation in the area of automatic prompt optimization will not only enhance the quality and plausibility of synthetic data but also support the shift toward Generalized AI.

%
%
%
\bibliographystyle{splncs04}

%

\end{document}